\documentclass[aps, prd, showpacs, nofootinbib, preprintnumbers,floatfix,letterpaper,longbibliography,superscriptaddress, reprint]{revtex4-1}

\usepackage[colorlinks=true, linkcolor=blue, citecolor=cyan]{hyperref}
\usepackage[utf8]{inputenc}
\usepackage{bm}
\usepackage{latexsym,amssymb,amsmath,amsfonts}
\usepackage{graphicx}
\usepackage{footnotebackref}
\usepackage{physics}
\usepackage[usenames]{xcolor}
\usepackage{soul}
\usepackage{upgreek}
\usepackage{subcaption}
\usepackage{makecell}
\usepackage{amssymb}
\usepackage{amsmath}
\usepackage{rotating}
\usepackage{color}
\usepackage{epsf}
\usepackage{fancyhdr}
\usepackage{ifthen}
\usepackage{slashed}
\usepackage{xspace}
\usepackage{tabularx}

\def\tildeSone{\tilde{s}_{1}}
\def\tildeStwo{\tilde{s}_{2}}
\def\d{\textrm{d}}
\def\hatOmega{{\hat{\Omega}}}
\def \hpx{{\tt HEALPix} }

\def\GW{\textrm{gw}}

\begin{document}

\title{Targeted search for the stochastic gravitational-wave background from the galactic millisecond pulsar population} 
\author{Deepali Agarwal}
\email{deepali@iucaa.in}
\affiliation{Inter-University Centre for Astronomy and Astrophysics (IUCAA), Pune 411007, India}

\author{Jishnu Suresh}
\email{jishnu.suresh@uclouvain.be}
\affiliation{Centre for Cosmology, Particle Physics and Phenomenology (CP3), Universit\'e catholique de Louvain, B-1348 Louvain-la-Neuve, Belgium} 

\author{Vuk Mandic}
\email{vuk@umn.edu}
\affiliation{University of Minnesota, Minneapolis, MN 55455, USA} 

\author{Andrew Matas}
\email{andrew.matas@aei.mpg.de}
\affiliation{Max Planck Institute for Gravitational Physics (Albert Einstein Institute), D-14476 Potsdam, Germany}

\author{Tania Regimbau}
\email{regimbau@lapp.in2p3.fr}
\affiliation{Laboratoire d’Annecy de Physique des Particules, CNRS, 9 Chemin de Bellevue, 74941 Annecy, France} 
\begin{abstract}
The millisecond pulsars, old-recycled objects spinning with high frequency $\mathcal{O}$ (kHz) sustaining the deformation from their spherical shape, may emit gravitational waves (GW). These are one of the potential candidates contributing to the anisotropic stochastic gravitational-wave background (SGWB) observable in the ground-based GW detectors. Here, we present the results from a likelihood-based targeted search for the SGWB due to millisecond pulsars in the Milky Way, by analyzing the data from the first three observing runs of Advanced LIGO and Advanced Virgo detector. We assume that the shape of SGWB power spectra and the sky distribution is known \textit{a priori} from the population synthesis model. The information of the ensemble source properties, i.e., the in-band number of pulsars, $N_{\textrm{obs}}$ and the averaged ellipticity, $\mu_\epsilon$ is encoded in the maximum likelihood statistic. We do not find significant evidence for the SGWB signal from the considered source population. The best Bayesian upper limit with $95\%$ confidence for the parameters are $N_{\textrm{obs}}\leq8.8\times10^{4}$ and $\mu_\epsilon\leq1.4\times10^{-6}$, which is comparable to the bounds on mean ellipticity with the GW observations of the individual pulsars. Finally, we show that for the plausible case of $N_{\textrm{obs}}=40,000$, with the one year of observations, the one-sigma sensitivity on $\mu_\epsilon$ might reach $1.5\times10^{-7}$ and $4.1\times10^{-8}$ for the second-generation detector network having A+ sensitivity and third-generation detector network, respectively.
\end{abstract}

\maketitle


\section{Introduction}

Dozens of gravitational-waves (GWs) sources have been cataloged~\cite{theligoscientificcollaboration2021gwtc21,theligoscientificcollaboration2021gwtc3,nitz20214ogc} using the data from the recently completed third observing run (O3) of Advanced LIGO \cite{aLIGO2015} and Advanced Virgo \cite{AdV_2014} detectors. These sources fall under the compact binary coalescence (CBC) category, particularly binary-black-hole mergers, binary-neutron-star mergers, and black-hole-neutron star mergers, whose signal lasts for seconds. The continuous GWs and stochastic gravitational-wave background (SGWB) are the interesting source categories yet to be detected. The continuous GWs is a persistent form of the gravitational radiation emitted at a nearly fixed frequency from a quadruple variation of matter, e.g., spinning neutron stars in isolated/binary system~\cite{universe7120474}. On the other hand, the SGWB is also a persistent signal but resultant of the incoherent superposition of GWs from a large number of sources with cosmological (e.g., inflationary GWs) and astrophysical origin (e.g., CBCs and neutron stars, etc.), and hence random in nature~\cite{Romano2017}. It is expected that the weak GW sources that are individually undetectable will produce a SGWB whose collective, incoherent signal will be detectable.

The SGWB can be categorized based on different angular distributions, i.e., isotropic and anisotropic or/and spectral distribution properties, i.e., broadband (with a power-law spectral model specific to source population) and narrowband sources. The astrophysical sources are also expected to produce anisotropic signal~\cite{anisoAGWB1,anisoAGWB2,anisoAGWB3,anisoAGWB4,anisoAGWB5,anisoAGWB6,anisoAGWB7,PhysRevLett.120.231101, Capurri:2021zli, PhysRevD.101.103513,Bellomo:2021mer,PhysRevD.101.081301,Regimbau:2022mdu}, and the upper limits placed on the estimator of SGWB amplitude~\cite{anisoAGWB1,DipongkarNS,DeLillo:2022blw} by isotropic searches could lead to conservative limits. The isotropic and directional searches were performed for the broadband SGWB combining the estimators from multiple frequency bins weighted by a power-law spectral model using the data from several runs of Advanced LIGO and Advanced Virgo detectors~\cite{O3Iso,O3BBR}.

In the past, a likelihood-based formalism was proposed and discussed in Refs.~\cite{EricSpH,DipongkarMultiBaseline,DipongkarNS}, to perform a targeted search for an extended anisotropic SGWB knowing \textit{a priori}, its angular distribution along with the spectral properties. This can improve the sensitivity of the search for the extended sources considerably. In this work, we adopt a similar formalism and perform a targeted search for the SGWB formed by the galactic millisecond pulsars population, using the data from the first three observing runs (O1, O2, and O3) of Advanced LIGO and Advanced Virgo observatories.

Out of $10^{8}-10^{9}$ neutron stars in the Milky Way galaxy~\cite{refId0}, $\sim40000$ recycled and rotation powered pulsars are expected to spin with period $<30$ms~\cite{lorimer_2012} called millisecond pulsars (MSPs). The MSPs with the asymmetric deformations around its spin axis (spinning with frequency $f$) may emit ``monochromatic" continuous GWs (at frequency 2$f$) in the frequency range of several 100 Hz to about 1 kHz where the ground-based GW detectors are sensitive\footnote{Several other mechanisms of the spinning neutron stars will also produce GWs, e.g., r-modes~\cite{lasky_2015}}. For a nonprecessing triaxial body with the spin axis along the z axis, the GW strain amplitude is proportional to the deformation parameter called ellipticity $\epsilon$ and is defined as
\begin{equation}\label{eq:def_epsilon}
    \epsilon = \frac{I_{xx}-I_{yy}}{I_{zz}}\,,
\end{equation}
where $I_{xx},I_{yy}$ and $I_{zz}$ are the principal moments of inertia (or $xx$, $yy$ and $zz$ component). In practice, the ellipticity is very small, i.e., $|\epsilon|\ll 1$ and $I_{xx}\simeq I_{yy} \simeq I_{zz} = I$. 

The neutron stars can serve as an astrophysical laboratory to probe the equation of states of matter at several nuclear saturation densities. The maximum deformation of the neutron star is a function of the equation of state, i.e., a stiffer equation of state allows larger deformations than the softer ones. Thus measuring the ellipticity can constrain the equation of state. The maximum ellipticity due to thermal pressure perturbation lies in the range $10^{-10}-10^{-7}$ for different chiral effective-field-theory
equation-of-state models~\cite{10.1093/mnras/stab2048}. \citet{PhysRevD.66.084025} claimed the ellipticity in range $\epsilon\sim10^{-9}-10^{-8}$ caused by the internal toroidal magnetic field for millisecond pulsars. Also, since the spin-down observed in the electromagnetic observations is due to GW emission, an average upper limit ellipticity of $\sim10^{-8}$ is calculated for millisecond pulsars called spin-down limit. The targeted search for GW signals from a nearby recycled pulsar (PSR J0711$-$6830) has bound ellipticity to be $\epsilon\leq 5.3\times10^{-9}$ which surpasses the indirect spin-down limit~\cite{https://doi.org/10.48550/arxiv.2111.13106}. We note that there is evidence for the existence of minimum ellipticity $\epsilon\geq 10^{-9}$ which indicates that GW radiation might be the dominant mechanism for the spin-down of MSPs~\cite{Woan_2018}.

Due to the weak signal strength, the individual detection of GWs produced by MSPs at a far distance (Galactic and extragalactic) may not be possible. However, these are potential candidates which contribute to the astrophysical SGWB~\cite{PhysRevD.84.083007,DipongkarMultiBaseline,PhysRevLett.122.081103,DipongkarNS,DeLillo:2022blw}. Hence, the SGWB searches can detect GWs from ensemble of MSPs and can give us more information about the MSPs' ensemble properties, like the number of MSPs within the search band (in-band number) and the average ellipticity.

The paper is structured as follows. In Sec.~\ref{sec:sgwb_methods}, we review the formalism for performing a cross-correlation-based targeted search for an anisotropic SGWB and derive a maximum likelihood statistic for the ``overall amplitude" of the SGWB. In Sec.~\ref{sec:MSP_sgwb}, we discuss the MSP population synthesis model, which we adopt to perform the stochastic search. We will also illustrate the method to prepare a \textit{template} for the spatial distribution. Details about the data and analysis pipeline are given in Sec.~\ref{sec:data_pipeline}. Following the analysis outlined in the previous section, we present the results from the search in Sec.~\ref{sec:results}. In Sec.~\ref{sec:future}, we will show the forecast on the expected sensitivity with the future detector network. We will conclude the article with the future prospects of the search in Sec.~\ref{sec:conclusions}.



\section{SGWB Search Methods}\label{sec:sgwb_methods}

Considering the GW strain data from two geographically separated detectors, the SGWB signal is expected to be correlated while the detector noise is uncorrelated. Hence, the searches~\cite{O3Iso,O3BBR} for SGWB are performed by constructing a cross-correlation spectral density (CSD) for a given baseline $\mathcal{I}$ (formed with the two detectors $1$ and $2$) as
    \begin{equation}\label{eq:data_CSD}
        C_{\mathcal{I}}(t;f)= \frac{2}{\tau}\,\tildeSone^{*}(t;f)\,\tildeStwo(t;f)\,,
    \end{equation}
where $\tildeSone$ and $\tildeStwo$ are the short-term Fourier transforms of the strain time series data of segment duration $\tau$ from detector $1$ and $2$ and centered around the time labeled by $t$. The expected value of $C_{\mathcal{I}}(t;f)$ is related to the one-sided power spectral density (PSD) $\mathcal{P}_{\hatOmega}(f)$ of the incoming GWs in the frequency range $f$ and $f+\d f$ per solid angle $\d^{2}\hatOmega$, if the source is in the direction $\hatOmega$, as
\begin{equation}\label{eq:expCSD}
        \langle C_{\mathcal{I}}(t;f)\rangle= \int \d^{2}\hatOmega \,\gamma^{\mathcal{I}}_{\hatOmega}(t;f)\,\mathcal{P}_{\hatOmega}(f)\,.
   \end{equation} 
Here, $\gamma^{\mathcal{I}}_{\hatOmega}$ denotes a detector-geometry dependent function, usually referred to as the overlap reduction function (ORF). The information of the detector response is encoded in this ORF, and it varies with the sidereal time, location of the detectors, and the frequency of the signal~\cite{Ballmer_2006,SanjitRadiometer,EricSpH,Romano2017}. We note that the observed CSD represents the signal from the collection of sources convolved with the detector response.

The source strain PSD $\mathcal{P}_{\hatOmega}(f)$ can be decomposed into the orthogonal bases $e_{\alpha}(\hatOmega)$, suitable to the angular distribution of the sources in the sky as
\begin{equation}
    \mathcal{P}_{\hatOmega}(f)= e_{\alpha}(\hatOmega)\,\mathcal{P}_\alpha(f)\,,
\end{equation}
using the Einstein sum convention. Depending on the source angular distribution, one can choose the basis function $e_{\alpha}(\hatOmega)$. The pixel basis $e_{\alpha}(\hatOmega)=\delta^{2}(\hatOmega-\hatOmega_\alpha)$ is the preferred choice for a point source where as the spherical harmonic basis $e_{\alpha}(\hatOmega)=Y_{l,m}(\hatOmega)$ is usually used for the extended source distributions. The unit of the elements $\mathcal{P}_\alpha(f)$ is Hz$^{-1}$sr$^{-1/2}$ in the spherical harmonic basis while Hz$^{-1}$ in the pixel basis. The analysis reported in this paper make use of the pixel basis to report the results. So, using Eq.~(\ref{eq:expCSD}) one can write the expected value of the CSD and the ORF, respectively, as
\begin{eqnarray}
        \langle C_{\mathcal{I}}(t;f)\rangle&=&\gamma^{\mathcal{I}}_{\alpha}(t;f)\,\mathcal{P}_{\alpha}(f)\,,\\
        \gamma^{I}_{\alpha}(t;f)&=&\int\d^{2}\hatOmega\,\gamma^{\mathcal{I}}_{\hatOmega}(t;f)\, e_{\alpha}(\hatOmega)\,.
   \end{eqnarray} 

In practice, we combine estimators from multiple time segments ($\sim80,000$), multiple baselines, and the frequency bins (when searching for broadband signal) to obtain a broadband ``average'' estimator of source strain PSD. In such cases, from both a central limit theorem and a \textit{weak signal limit}, the CSD is expected to follow a Gaussian distribution with variance $P_1(t;f)P_2(t;f)$~\cite{AinFolding}. Here $P(t,f)$ is the one-sided noise PSD for the individual detector. Now one can write the combined likelihood $L$ for the CSD as
\begin{eqnarray}\label{eq:lkhdCSD}
   L\propto\prod_{\mathcal{I}\,t,f}\,\mathrm{exp}\,\bigg[-\frac{1}{2}\,\bigg(C_{\mathcal{I}}(t;f)-\gamma^{\mathcal{I}}_{\alpha}(t;f)\,\mathcal{P}_{\alpha}(f)\bigg)^{*}\nonumber
   \\\frac{1}{P_1(t;f)P_2(t;f)} \bigg(C_{\mathcal{I}}(t;f)-\gamma^{\mathcal{I}}_{\alpha'}(t;f)\,\mathcal{P}_{\alpha'}(f)\bigg)\bigg]\,.
\end{eqnarray}
If, we further decompose the source strain PSD in terms of a frequency dependent factor $\bar{H}_f$, a direction dependent factor $\hat{\mathcal{P}}_\alpha$ and an ``overall amplitude" denoted by a scalar $A$, then
\begin{equation}
    \mathcal{P}_{\alpha}(f) = A\,\bar{H}_f\, \hat{\mathcal{P}}_\alpha\,.
\end{equation}
Detailed discussions on the astrophysical origin of the quantities in the right hand side of the above equation are given in Sec.~\ref{sec:MSP_sgwb}. Assuming that the $\bar{H}_f$ and $\hat{\mathcal{P}}_\alpha$ are confidently known, then maximum likelihood (ML) estimator of $A$ and its mean $\langle A\rangle$ are given as~\cite{EricSpH,DipongkarMultiBaseline} 
\begin{equation}\label{eq:hatA}
 \hat{A}=\frac{\mathbf{X}^\dagger\bm{\hat{\mathcal{P}}}}{\bm{\hat{\mathcal{P}}}^\dagger\mathbf{\Gamma}\bm{\hat{\mathcal{P}}}}\quad;\quad\langle \hat{A}\rangle=A\,.
\end{equation}
Here $\mathbf{X}$ is the ``dirty map''
\begin{equation}\label{eq:dirty_map}
    \mathbf{X}\equiv X_\alpha=\sum_{\mathcal{I}\,t,f} \,\frac{C_{\mathcal{I}}(t;f)\,\gamma^{\mathcal{I}*}_{\alpha}(t;f)\,\bar{H}_f}{P_1(t;f)P_2(t;f)}\,,
\end{equation}
and $\mathbf{\Gamma}$ represents the Fisher information matrix~\cite{SanjitRadiometer,EricSpH},
\begin{equation}\label{eq:fisher}
    \mathbf{\Gamma}\equiv \Gamma_{\alpha,\alpha'}=\sum_{\mathcal{I}\,t,f} \,\frac{\gamma^{\mathcal{I}*}_{\alpha}(t;f)\,\gamma^{\mathcal{I}}_{\alpha'}(t;f)\,\bar{H}^2_f}{P_1(t;f)P_2(t;f)}\,.
\end{equation}
Following these definitions, it is straightforward to write the likelihood in Eq.~(\ref{eq:lkhdCSD}) in terms of the dirty map and Fisher information matrix as
\begin{equation}\label{eq:lkhdDirty}
   L\propto\mathrm{exp}\,\bigg[-\frac{1}{2}\bigg(\mathbf{X}-A\,\mathbf{\Gamma}.\bm{\hat{\mathcal{P}}}\bigg)^{\dagger}\mathbf{\Gamma^{-1}}\bigg(\mathbf{X}-A\,\mathbf{\Gamma}.\bm{\hat{\mathcal{P}}}\bigg)\bigg]\,.
\end{equation}

It is interesting to note that the ML estimator, $\hat{A}$ has a similar form to the matched-filter statistic used in CBC searches~\cite{PhysRevD.85.122006}. In our analysis, we are essentially ``matching'' the observed dirty map with the model describing the sky distribution of the source power. The model is obtained by convolving the \textit{template} $\bm{\hat{\mathcal{P}}}$, sky distribution known a priori, with the detector response (i.e., the forward modelling). From the likelihood, the variance and signal-to-noise ratio (SNR) of $\hat{A}$ are given as~\cite{EricSpH,DipongkarMultiBaseline}
\begin{eqnarray}
 \sigma^2_{\hat{A}}&=&\frac{1}{\bm{\hat{\mathcal{P}}}^\dagger\mathbf{\Gamma}\bm{\hat{\mathcal{P}}}}\label{eq:sigmaHatA}\,,\\
 \rho_{\hat{A}}&=&\frac{\mathbf{X}^\dagger\bm{\hat{\mathcal{P}}}}{\sqrt{\bm{\hat{\mathcal{P}}}^\dagger\mathbf{\Gamma}\bm{\hat{\mathcal{P}}}}}\label{eq:SNRA}\,.
\end{eqnarray}
The Gaussian nature of the CSD makes sure that the dirty map $\mathbf{X}$, the ML estimator $\hat{A}$ of the amplitude $A$, and its SNR $\rho_{\hat{A}}$ follow the Gaussian distribution with their mean and variance. These properties will be useful in assigning the significance to the observed data and setting the upper limit on the source parameters in case of no detection.

It is also interesting that if the template is a vector with a single nonzero element, having a value equal to 1, then $\hat{A}$ is equivalent to the broadband radiometer search estimator, a measure of the strength of the GWs signal from a pixel or $(l,m)$ mode in the sky~\cite{O3BBR}. On the other hand, if the template is a vector with the elements having a value of 1, then $\hat{A}$ is identical to the isotropic search estimator~\cite{O3Iso}.

In the usual map-making process, ``clean map'' is the ML estimator of the ``true'' sky map $\bm{\hat{\mathcal{P}}}$ obtained through deconvolution process~\cite{EricSpH,SanjitRadiometer,Sambit,O1O2Folded}. It is also possible to rewrite the likelihood given in Eq.~(\ref{eq:lkhdDirty}) to obtain the ML estimator $\hat{A}$ of the amplitude in terms of this clean map~\cite{DipongkarMultiBaseline}. However, the deconvolution involves the inversion of the highly ill-conditioned Fisher information matrix. Due to the insensitivity of the detector to certain modes/directions in the sky, the inversion of the Fisher matrix leads to the inverted noise boost~\cite{EricSpH,SanjitRadiometer,renzini1,renzini2,O1O2Folded}, thus making the deconvolution process a challenge. We choose to work with the dirty map and avoid unnecessary information loss and numerical errors that may arise in the regularization process.


\begin{figure*}
    \centering
     \begin{subfigure}[h]{0.45\textwidth}
         \centering
         \includegraphics[height = 0.45\textwidth]{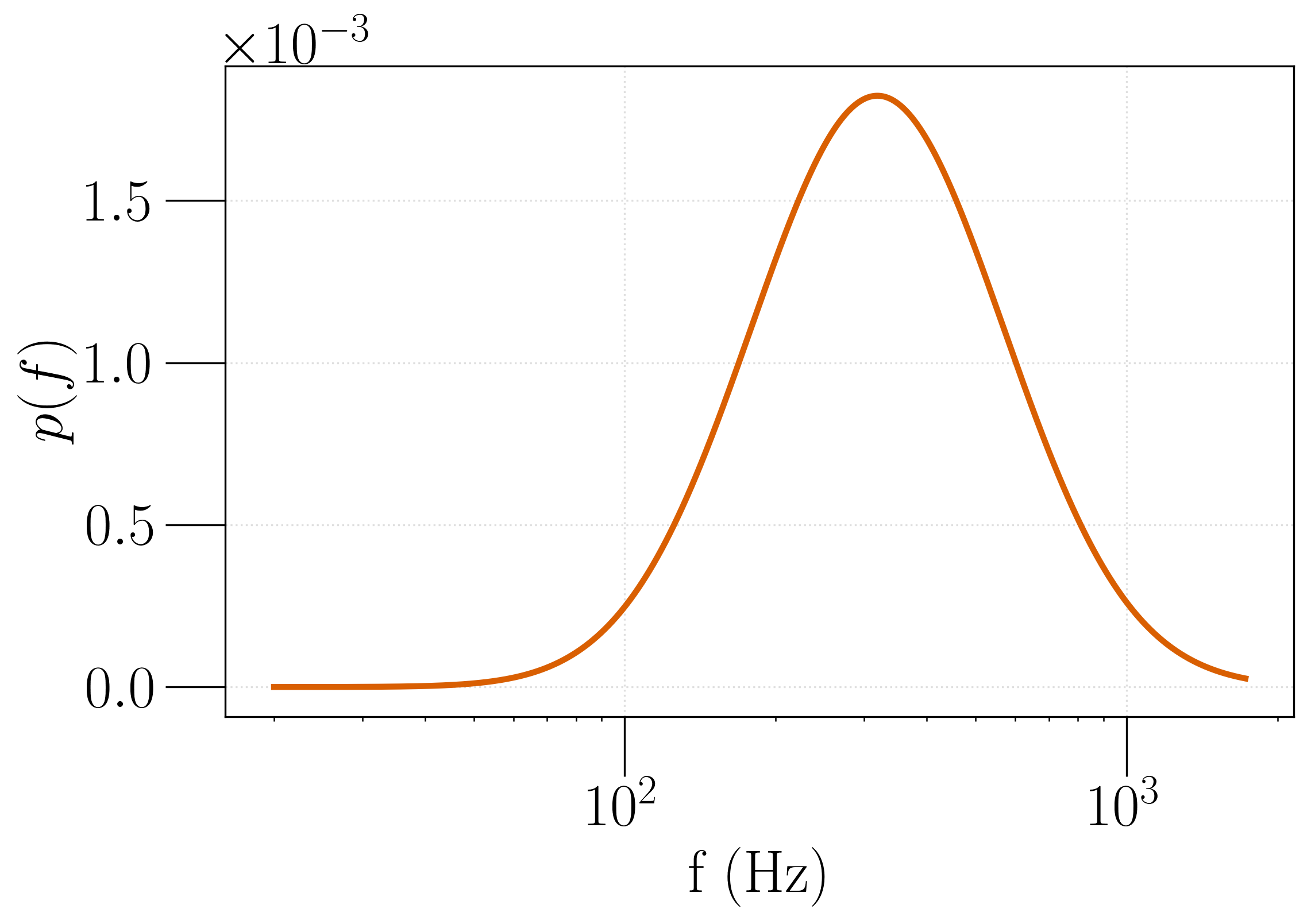}
         \caption{}
         \label{fig:Nf}
     \end{subfigure}
     \hspace{0.2cm}
     \begin{subfigure}[h]{0.45\textwidth}
         \centering
         \includegraphics[height = 0.45\textwidth]{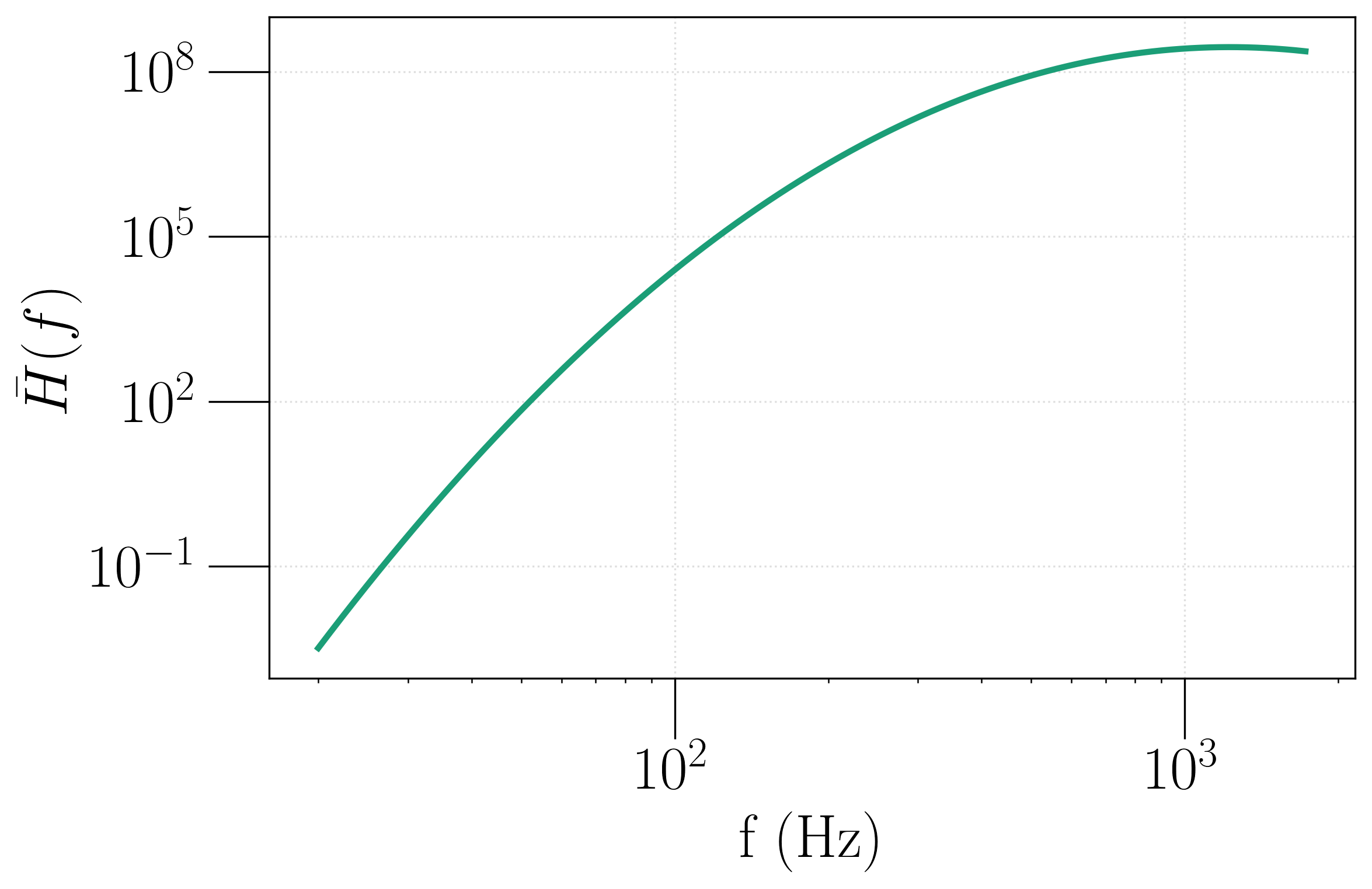}
         \caption{}
         \label{fig:Hf}
     \end{subfigure}\\
     \begin{subfigure}[h]{0.45\textwidth}
         \centering
         \includegraphics[height = 0.45\textwidth]{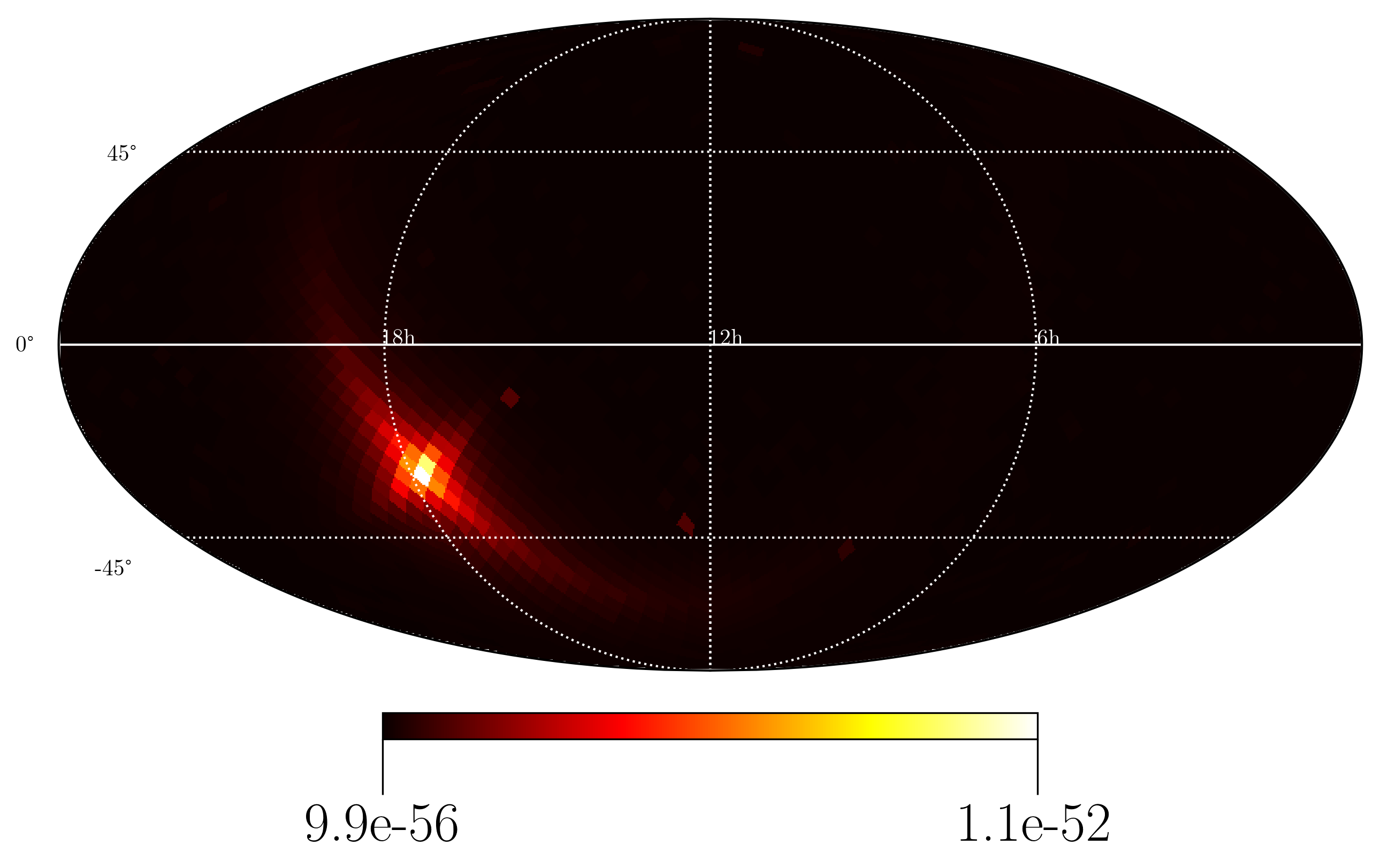}
         \caption{}
         \label{fig:templateExponential}
     \end{subfigure}
     \hspace{0.2cm}
     \begin{subfigure}[h]{0.45\textwidth}
         \centering
         \includegraphics[height = 0.45\textwidth]{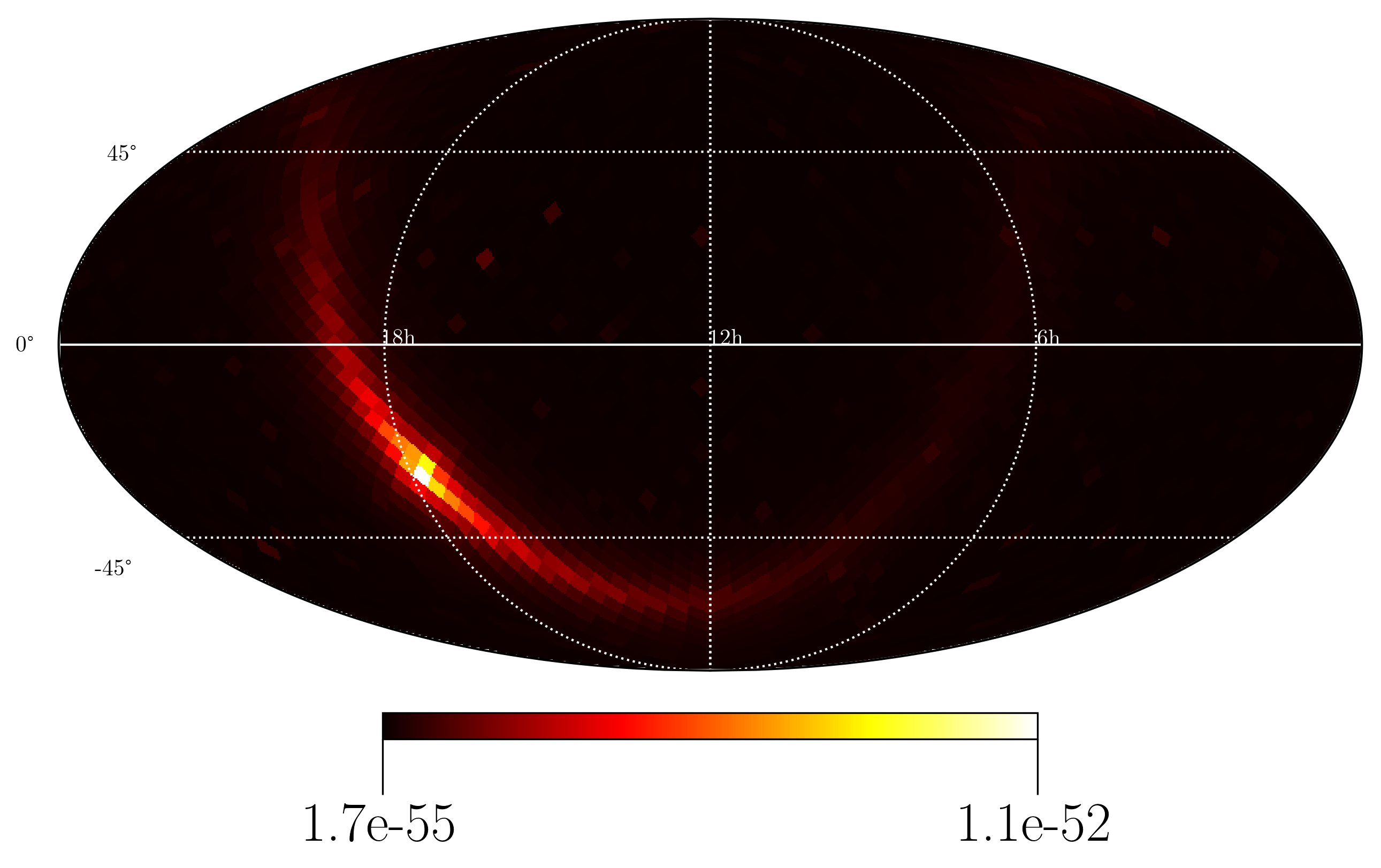}
         \caption{}
         \label{fig:templateGaussian}
     \end{subfigure}
        \caption{The MSP population synthesis model used in the analysis. (a) The probability density profile, $p(f)$ as a function of the GW frequency $f$. (b) The expected spectral dependence $\bar{H}(f)$ of SGWB signal. The maps (c) and (d) represent the template $\mathcal{\hat{P}}$ for the spatial distribution of the source power with exponential and Gaussian density profiles for the radii, respectively.}
        \label{fig:Templates}
\end{figure*}

\section{MSP Population Synthesis Model}\label{sec:MSP_sgwb}
In this section, we will be discussing the expected PSD of the MSP population. The strain PSD for SGWB signal from the neutron star population emitting GWs in the frequency range $f$ to $f+\d f$ and lying in solid angle $\hatOmega$ and $\hatOmega+\d^2 \hatOmega$ is given by (see Appendix~\ref{sec:PSD_derivation})
 \begin{equation}\label{eq:MSP_PSD}
     \mathcal{P}{(f,\hatOmega)}=\underbrace{N_{\textrm{obs}}\langle \epsilon^2 \rangle_{s}}_{A}\,\underbrace{f^4\,p(f)}_{\bar{H}_f}\, \frac{32\,\pi^4\,G^2\,\langle I^2 \rangle_{s}}{5c^8}\,\langle r^{-2}\rangle_{s}\,p(\hatOmega)\,.
 \end{equation}
Since we are working in the pixel basis, the elements of the template $\mathcal{\hat{P}}$ are given by
\begin{equation}\label{eq:template}
    \mathcal{\hat{P}}_\alpha=\int_{\hatOmega_\alpha}^{\hatOmega_\alpha+d{\hatOmega_\alpha}}\frac{32\,\pi^4\,G^2\,\langle I^2 \rangle_{s}}{5c^8}\,\langle r^{-2}\rangle_{s}\,p(\hatOmega)\,d\hatOmega
\end{equation}
Above $N_{\textrm{obs}}$ is the total number of neutron stars within the observing band and sky patch. Therefore,
 \begin{equation}
     N_{\textrm{obs}} = \int_{f_{\mathrm{min}}}^{f_{\mathrm{max}}} \d f\,p(f) \int_{\mathrm{sky}} \d \hatOmega \,\pi_{\hatOmega} \int_{0}^{\infty}\d r\,r^{-2}p(r)\,,
 \end{equation}
where $f_{\mathrm{min}}$ and $f_{\mathrm{max}}$ are, respectively, the lower and upper limits of the observed frequency band. The probability density of a MSP to be observed in the direction $\hatOmega$ at distance $r$ from the Earth and emitting GWs at frequency $f$ are encoded in $p(\hatOmega)$, $p(r)$, and $p(f)$. The parameters $I$ and $\epsilon$ are the principle moment of inertia and deformation parameter called ellipticity [Eq.~(\ref{eq:def_epsilon})] of each neutron star. The $\langle...\rangle_{s}$ denotes the ensemble average over the source population. Subscript ``s'' in the angular bracket is introduced to distinguish it from the ensemble average over noise in Eq.~(\ref{eq:hatA}). In this study, we will be using a fiducial value of $\langle I^2 \rangle_{s} =(1.1\times10^{38}\,\text{kgm}^2)^2$, which is constrained very well from nuclear physics studies~\cite{worley2008nuclear}.
 
In order to describe the SGWB signal from the Galactic MSPs, a model of their spatial and frequency distribution is required. The determination of the intrinsic distribution of the spin-period, magnetic field, period derivative, and spatial coordinates for the MSPs is an ambitious goal in the field of electromagnetic astronomy as well. There have been studies to understand the underlying distribution based on the statistical analyses of artificial MSPs that pass the criteria for detection and comparing them with the detected MSPs~\cite{Story_2007,lorimer_2012,Lorimer2015,FremiFit2013}. Next, we will discuss the model adopted for the spectral shape and the template for the spatial distribution constrained by electromagnetic observations.

 \begin{table*}[t]
    \centering
    \begin{tabular}{c|c|c|c|c}
        \hline\hline
        \multicolumn{5}{c}{\textbf{O1+O2+O3 results}} \\
        \hline
       \hline
       \multicolumn{1}{c|}{Baseline}& \multicolumn{2}{c|}{Exponential radial distribution} & \multicolumn{2}{c}{Gaussian radial distribution}\\
        \hline
        \multicolumn{1}{c|}{}&~ $(\hat{A}\pm\sigma_{\hat{A}})\times10^{-8}$~ & ~$\rho_{\hat{A}}\,(p-\textrm{value} \%)$ ~ &~ $(\hat{A}\pm\sigma_{\hat{A}})\times10^{-8}$~ &~ $\rho_{\hat{A}}\,(p-\textrm{value} \%)$~\\
         \hline
          O3-HL & 5.7 $\pm$  6.2 & 0.92 (18) & 3.4 $\pm$ 6.3  & 0.54 (30)  \\
         \hline
         O3-HV  &   120 $\pm$  53 & 2.3 (1.2) & 96 $\pm$ 44 & 2.2 (1.4)\\
         \hline
          O3-LV &   17 $\pm$  31 & 0.54 (29) & 59 $\pm$ 29 & 2.0 (2.1) \\
          \hline
          O2-HL & -17 $\pm$  24 & -0.69 (76) & -8.8 $\pm$ 25  & -0.36 (64)  \\
          \hline
          O1-HL & -54 $\pm$  51 & -1.1 (85) & -56 $\pm$ 53  & -1.1 (86)  \\
          \hline
        \textbf{O1+O2+O3} &   $\mathbf{5.4 \pm  5.8}$ & \textbf{0.92 (18)} & $\mathbf{5.9 \pm 5.9}$ & \textbf{1.0 (16)} \\
         \hline\hline
    \end{tabular}
    \caption{Here, we report the results of the targeted stochastic search analysis using the data from the first three observing runs of Advanced LIGO (H and L) and Advanced Virgo (V) detectors, hence the five individual datasets, i.e., O3-HL, O3-LV, O3-HV, O2-HL, and O1-HL and with the combined network, O1+O2+O3. The observed overall amplitude, $\hat{A}$ with the uncertainty, $\sigma_{\hat A}$ and SNR, $\rho_{\hat A}$ are obtained using the two templates for the spatial distribution created for the exponential and Gaussian distributed radial coordinate. The results are assessed through the $p$-value against the null hypothesis, which is that the data is pure Gaussian noise. We do not claim any detection since the obtained $p$-values do not pass the threshold, $5\%$.}
    \label{tab:O3_Results}
\end{table*}
 
 \subsection{Frequency Dependence Model}
The likelihood analysis of a sample of $\sim56$ radio MSPs observed in the ``first generation'' of Parkes multibeam surveys~\cite{Parkes1,Parkes2,Parkes3,Parkes4,Parkes5,Parkes6} found that the underlying distribution of spin period of MSPs can be best fitted with log-normal function form. Interestingly, these findings are consistent with the current (large) sample of $\sim206$ MSPs within $95\%$ credible region~\cite{Lorimer2015}. For our analysis, we consider the best fit parameter values, as given in \citet{Lorimer2015}. We also modify the probability density function (PDF) by changing the spin-period $P$ to the GW frequency variable $f=2/P$ (Hz) to well suit the analysis described in this paper. The modified PDF is given as
\begin{equation}\label{eq:freq_density}
     p(f|\mu,\sigma)=\frac{1}{\sqrt{2\pi}\,f\,\sigma}\mbox{exp}\left[ -\frac{(\mbox{ln}(f)-\mu)^2}{2\sigma^2}\right]\,,
\end{equation}
where mean $\mu=6.1$ and variance $\sigma=0.58$. The probability density profile for the GW frequency and the spectral shape of SGWB using Eq.~(\ref{eq:MSP_PSD}) are illustrated in Figs.~\ref{fig:Nf} and~\ref{fig:Hf}, respectively. Note that the peak of PDF at $\sim400$ Hz is disappeared in the figure showing the spectral shape of SGWB, since the luminosity of individual MSP scales as $f^6$ [see Eq.~(\ref{eq:MSP_power})].
 
 \subsection{Spatial Distribution Model}\label{ssec:Spatial_distribution}
One can write the radial and height distribution for the MSP population in terms of the exponential functions as
\begin{equation}\label{eq:pdfRadial_exp}
     p(R,z)\propto \mbox{exp}(-R/R_0)\,\mbox{exp}(-|z|/z_0)\,,
 \end{equation}
where $R$ and $z$ (having unit kpc in the galactocentric coordinate system) define the pulsar's distance from the galactic center and the height of the pulsar above the galactic plane. We use the best-fit values for the parameters $R_0\approx4$ kpc and $z_0\approx1$ kpc as given in~\citet{FremiFit2013}. The polar axis passes through the galactic center perpendicular to the galactic plane. The PDF for polar angle $\phi$ can be assumed to be uniform between $[0,2\pi]$ with $\phi=0$ measured along the axis connecting the galactic center to Earth. 
On the other hand, other models for the radial distribution of MSPs are also explored in the literature~\cite{lorimer_2012} by considering a half-Gaussian distribution function as
 \begin{equation}\label{eq:pdfRadial_Gauss}
     p(R,z)\propto \mbox{exp}(-R^2/2\Sigma^2_r)\,\mbox{exp}(-|z|/z_0)\,,
 \end{equation}
where the radial and vertical heights are constrained to be $\Sigma_r=7.5$ kpc and $z_0\approx0.5$ kpc from the statistical analysis of the observed MSPs in radio band along with the uniform distribution for the polar angle. In this work, we will analyze the data considering both the PDFs for radial coordinate. The template or model map $\hat{\mathcal{P}}$ is obtained by following the steps given below:
\begin{enumerate}
    \item We draw random locations of $N=10^5$ pulsars with the density function given above in terms of galactocentric coordinates $\{R,z,\phi\}$.
    \item We then convert the pulsar positions from galactocentric coordinates $\{R,z,\phi\}$ to equatorial coordinates $\{\mathrm{RA},\mathrm{Dec},r\}$. Here, we assumed the sun to be at 8.12 kpc away from the galactic center~\cite{galcen_distance} and at 20.8 pc height~\cite{10.1093/mnras/sty2813} above the Galactic plane.
    \item Next the simulated pulsars are binned into 3072 \hpx pixels with $\text{n}_{\text{side}}=16$~\cite{HEALPix,Zonca}. We then calculate $r^{-2}$ for each pulsar and compute the average over that for each pixels (see Eq.~(\ref{eq:OmegaSource}) for a detailed derivation).
\end{enumerate}
 
The final step described above gives us the map of $\langle r^{-2}\rangle\,p(\hatOmega)$. We multiply it with the constant [see Eq.~(\ref{eq:template})] to obtain $\hat{\mathcal{P}}$. We then create 1000 such realizations following the above three steps recursively. The average of these realizations is considered as the \textit{template} for the MSP population. It is worth mentioning that, by considering the mean of these realizations, one can suppress the large power (statistical fluctuations) from the pulsars outside of the Galactic plane. The templates for both the exponential distribution and the Gaussian distribution of the radial coordinate are shown in Figs.~\ref{fig:templateExponential} and~\ref{fig:templateGaussian}, respectively.

Given these population properties, we are interested in constraining the ensemble properties of MSPs, i.e., $N_{\textrm{obs}}$ and averaged ellipticity $\mu_\epsilon$. However, the estimator $A$ of our search has information on the average of squared ellipticity, $\langle\epsilon^2\rangle_{s}$ [Eq.~(\ref{eq:MSP_PSD})] which is related to the $\mu_\epsilon$ as
 \begin{equation}
     \langle\epsilon^2\rangle_{s}=\mu_\epsilon^2+\Sigma_\epsilon^2\,.
 \end{equation}
Estimating intrinsic variance $\Sigma_\epsilon$ of ellipticity requires its distribution to be known. However, the actual distribution is not confidently known. Thus we assume,
\begin{equation}
    \langle\epsilon^2\rangle_{s}=\mu_\epsilon^2 \implies \, A=N_{\textrm{obs}}\,\mu_\epsilon^2\,.
\end{equation}
The above approximation is valid if the intrinsic variance is small compared to the averaged ellipticity, i.e., $\Sigma_\epsilon\ll\mu_\epsilon$. Even though this leads to bias in the estimator, in the rest of the paper, we assume this approximation is valid~\cite{DipongkarNS,DeLillo:2022blw}.
 


\section{Data and pipeline}~\label{sec:data_pipeline}
For the analysis, we use the data from the first three observing runs (O1, O2, and O3) of Advanced LIGO's Hanford (H) and Livingston (L) and Advanced Virgo (V) detectors calibrated by LIGO-Virgo-KAGRA collaboration~\cite{Sun_2020,https://doi.org/10.48550/arxiv.2107.00129,Acernese_2022,Davis_2021}. The data is now available publicly~\cite{RICHABBOTT2021100658,o3_data}. The strain time-series data is processed in a similar way as in ~\citet{O1paper,O2paper,O3BBR} to obtain the CSDs for individual datasets/baselines, i.e., O1-HL, O2-HL, and HL, LV, and HV in O3 run, as well as the PSDs for individual detectors. These quantities are computed for the segments of $\tau=192$ s long duration and 1/32 Hz frequency resolution along with the observing band of 20-1726 Hz [see Eq.~(\ref{eq:data_CSD})]. The data quality cuts in the time domain and the frequency domain to remove the non-Gaussian features and the known artifacts are applied identically as in~\citet{O1paper,O2paper,O3BBR}. The CSDs and PSDs are further compressed to one sidereal day using the folding algorithm~\cite{AinFolding,ligo_scientific_collaboration_virgo_coll_2022_6326656}. In the next step, we prepare the dirty map $\mathbf{X}$ [Eq.~(\ref{eq:dirty_map})] and the Fisher information matrix $\mathbf{\Gamma}$ [Eq.~(\ref{eq:fisher})] for each baseline with the folded data and {\tt PyStoch} pipeline~\cite{AinPystoch} in \hpx grid of $3072$ pixels in pixel basis. The dirty map and the Fisher information matrix for the combined network (O1+O2+O3) can be obtained by combining the estimators from individual baselines/observing runs using Eqs.~(\ref{eq:dirty_map}) and~(\ref{eq:fisher}).



\section{Observational Results}\label{sec:results}

With the observed data, we estimate the overall amplitude $\hat{A}$ [Eq.~(\ref{eq:hatA})] and its SNR $\rho_{\hat{A}}$ [Eqs.~(\ref{eq:sigmaHatA}) and~(\ref{eq:SNRA})], using the prepared dirty map and Fisher matrix for individual datasets and combined network (O1+O2+O3) along with the prepared templates $\hat{\mathcal{P}}$ as detailed in Sec.~\ref{ssec:Spatial_distribution}. The results of the analysis are obtained in two steps. First, the observed data is assessed against the null hypothesis by assigning the $p$-values. In the second step, we determine the $90\%$ confidence credible intervals (along with $95\%$ confidence upper limits) for the parameters defining the ensemble properties of the MSPs population; specifically, the in-band number of MSPs, $N_{\textrm{obs}}$ and averaged ellipticity, $\mu_\epsilon$.

\subsection{Significance}
To compute the $p$-value, we use the statistical property of the observed SNR of the overall amplitude $\rho_{\hat{A}}$ that it is a normal distributed random variable with zero mean and standard deviation 1 [see Eqs.~(\ref{eq:lkhdDirty}) and~(\ref{eq:SNRA})] in the absence of a signal. The results are summarised in Table~\ref{tab:O3_Results}. The observed SNR from the O1+O2+O3 dataset is $\rho_{\hat{A}}=0.92$ with $p$-value=18$\%$, if the exponential density profile for the radial coordinate is considered. On the other hand, using the template with the Gaussian distributed radial coordinate, the observed SNR from the O1+O2+O3 network is $\rho_{\hat{A}}=1.0$ with $p$-value=16$\%$. The observed SNR is consistent with the Gaussian noise. Hence, the results conclude that we do not find significant evidence for the SGWB from the galactic millisecond pulsars. We also note that the current observational data is not sensitive enough to distinguish between the spatial distribution models.

\begin{figure*}
     \centering
     \begin{subfigure}[h]{0.4\textwidth}
         \centering
         \includegraphics[width=\textwidth]{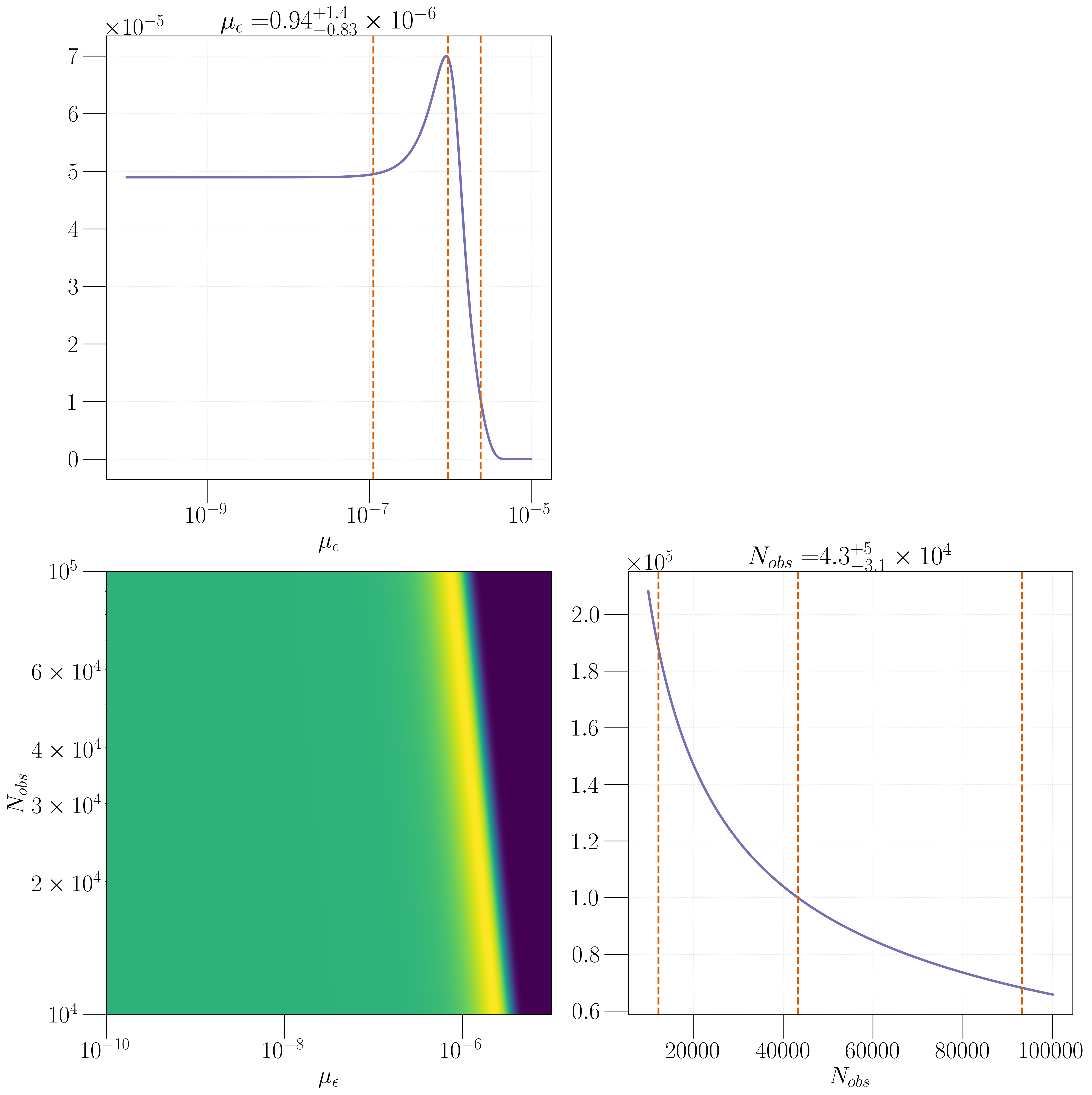}
         \caption{}
         \label{fig:Exp_uniform}
     \end{subfigure}
     \hspace{1cm}
     \begin{subfigure}[h]{0.4\textwidth}
         \centering
         \includegraphics[width=\textwidth]{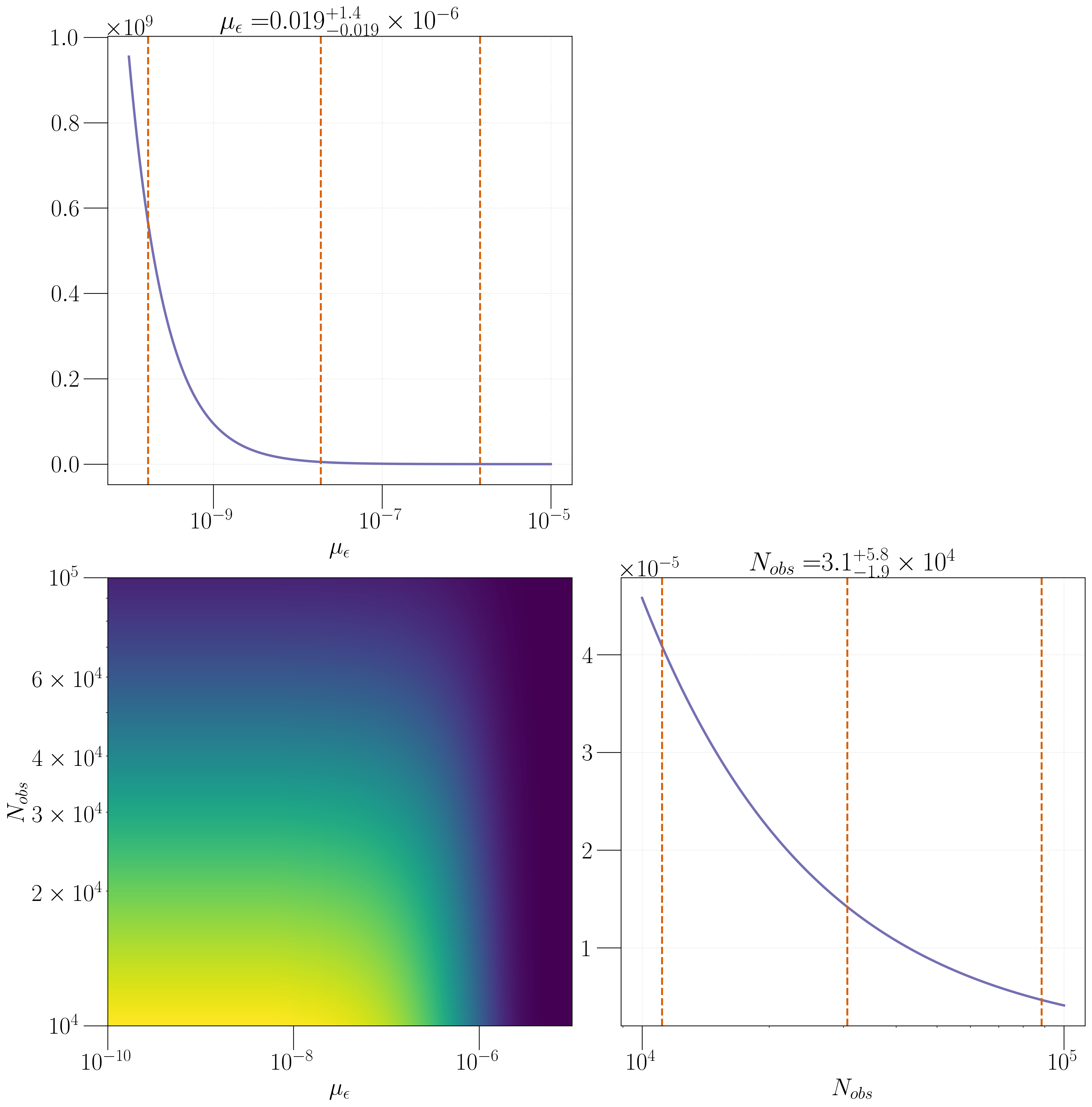}
         \caption{}
         \label{fig:Exp_logUniform}
     \end{subfigure}\\
      \vspace{0.2cm}
     \begin{subfigure}[h]{0.4\textwidth}
         \centering
         \includegraphics[width=\textwidth]{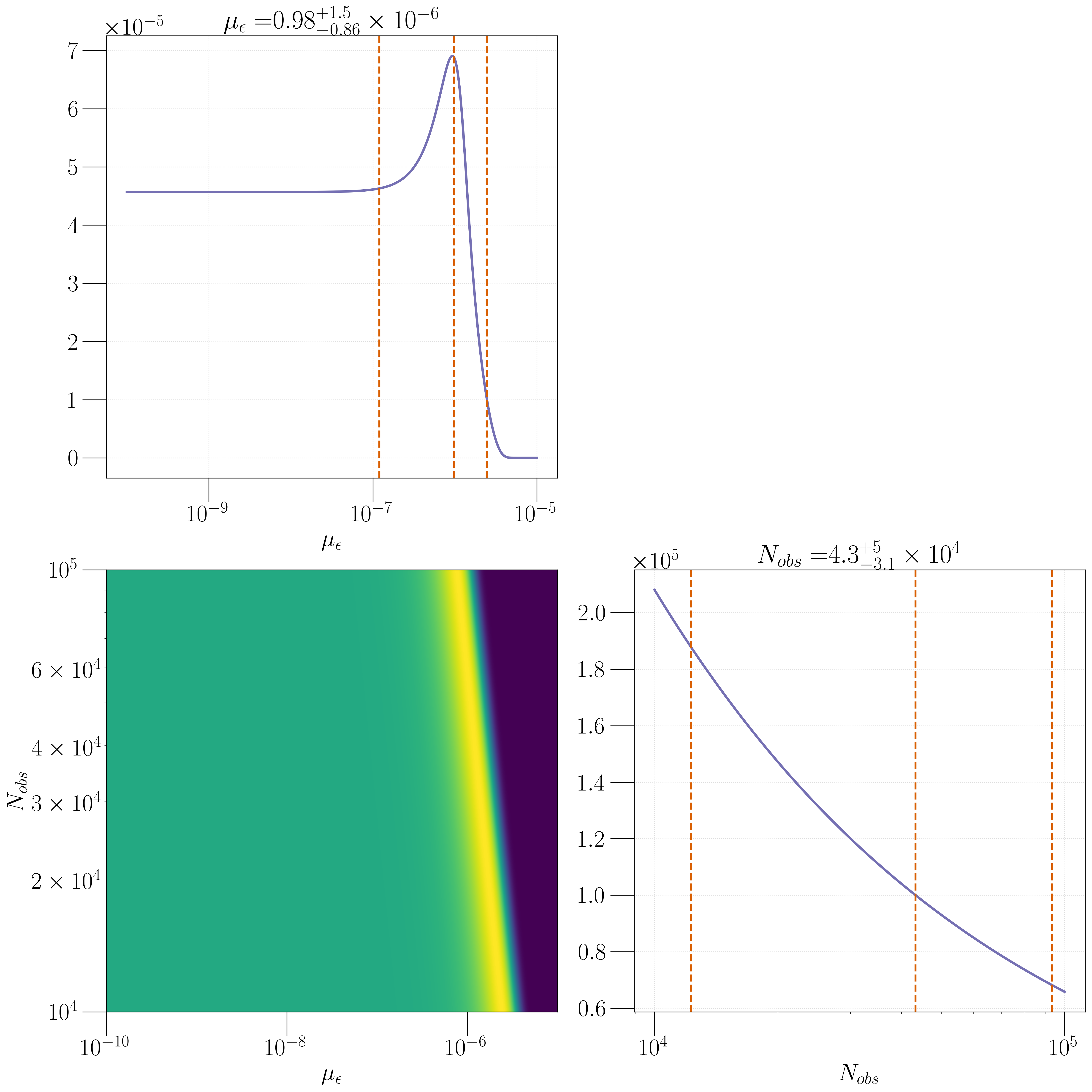}
         \caption{}
         \label{fig:Gauss_Uniform}
     \end{subfigure}
     \hspace{1cm}
     \begin{subfigure}[h]{0.4\textwidth}
         \centering
         \includegraphics[width=\textwidth]{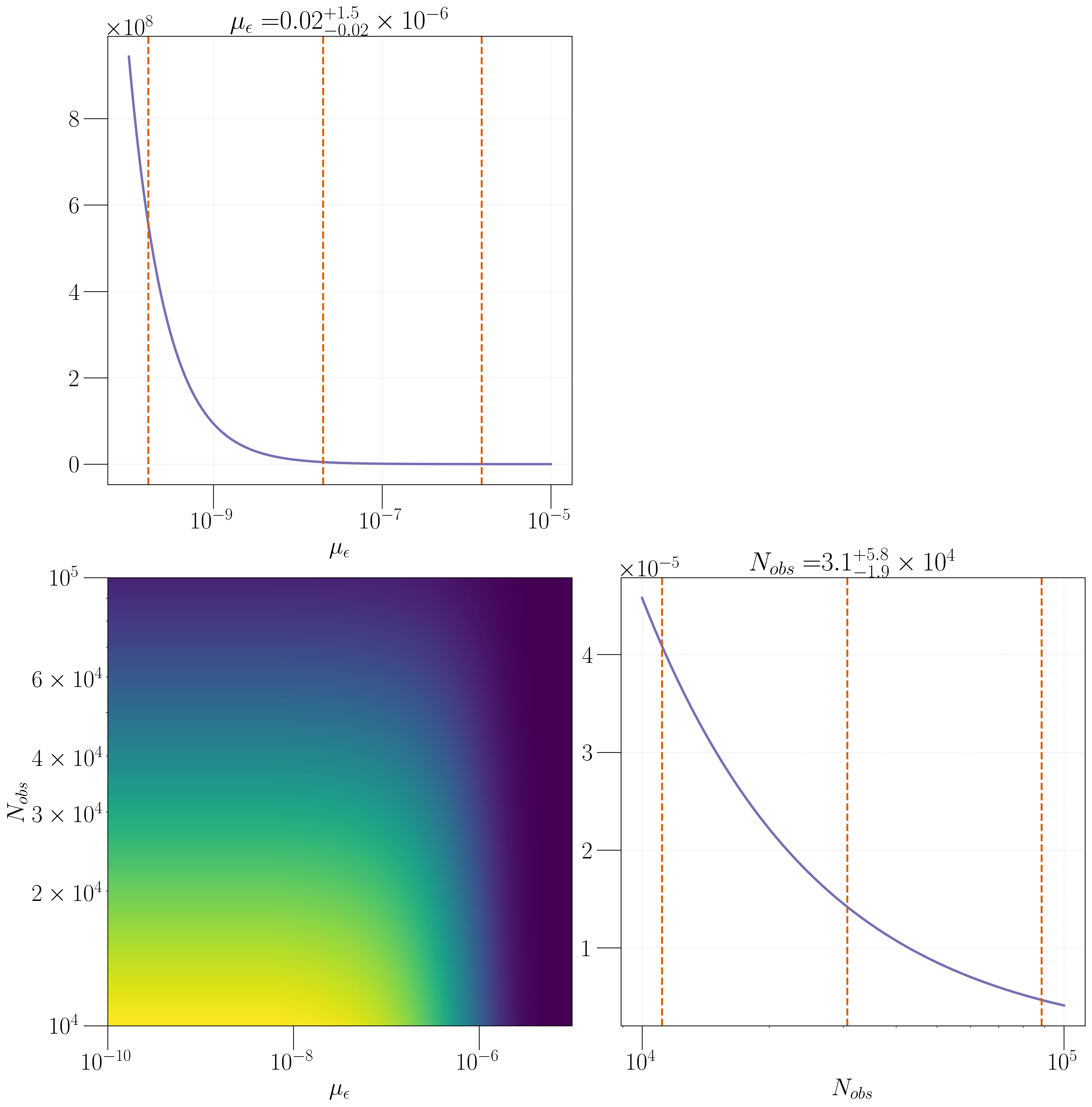}
         \caption{}
         \label{fig:Gauss_logUniform}
     \end{subfigure}
        \caption{The corner plot depicts the joint and marginalized posterior density for the source parameters ($N_{\textrm{obs}}$ and $\mu_\epsilon$) using the O1+O2+O3 data. The prior for the parameters are uniform (left) or log-uniform (right) over parameter ranges $\mu_\epsilon\in[10^{-10},10^{-5}]$ and $N_{\textrm{obs}}\in[10^{4},10^{5}]$. The dashed lines (and title) in the 1-D plot show the median values along with the (0.05,0.95) quantile values. The caption for each subplot shows the combination of radial density profile and assumed priors. (a) Exponential distribution and uniform prior (b) Exponential distribution and log-uniform prior (c) Gaussian distribution and uniform prior (d) Gaussian distribution and log-uniform prior. }
        \label{fig:O3_PE}
\end{figure*}


\subsection{Constraining The Source Parameters}

The ensemble properties of the MSP population are inferred using the Bayesian analysis. As we discussed in the previous sections, the PDF for the observed overall amplitude $\hat{A}$ can be assumed to be a Gaussian distribution with mean $A=N_{\textrm{obs}}\mu_\epsilon^2$ and standard deviation $\sigma_{\hat{A}}$. The two sets of prior are considered: uniform and log-uniform prior distributions for $N_{\textrm{obs}}$ and $\mu_\epsilon$ over the ranges $[10^4,10^5]$ and $[10^{-10},10^{-5}]$ respectively. Second, a log-uniform distribution considering the same maximum and minimum range for both the parameters. Given the likelihood and priors, the joint and marginalized posterior densities are computed numerically. The joint and marginalized posterior densities along with the median and $90\%$ credible interval for the parameters are shown in Fig.~\ref{fig:O3_PE}. In this figure, we show four combinations, i.e., uniform and log uniform prior for the parameters along with the observed $\hat{A}$ using the O1+O2+O3 network for the Exponential and Gaussian density profiles. 
The best $95\%$ confidence upper limits on the source parameters are obtained using the log-uniform prior: they are $\mu_\epsilon\leq1.4\times10^{-6}$ and $N_{\textrm{obs}}\leq8.8\times10^{4}$. The limit on the averaged ellipticity $\mu_\epsilon$ is consistent with the predicted minimum ellipticity of $\geq10^{-9}$~\cite{Woan_2018}. 
 

 \begin{figure}
     \centering
         \includegraphics[width=0.45\textwidth]{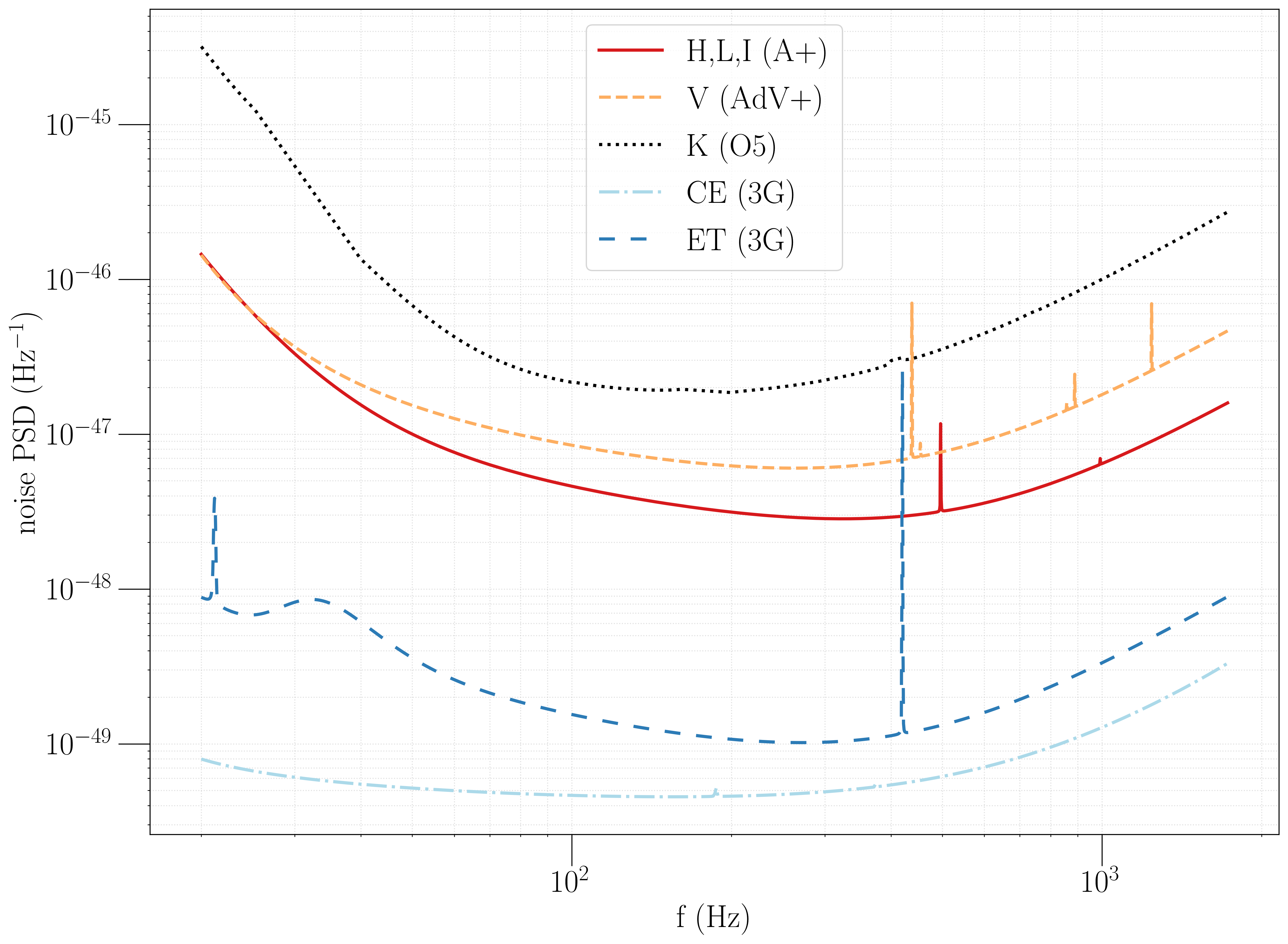}\\\hspace{0.1cm}
         \includegraphics[width=0.45\textwidth]{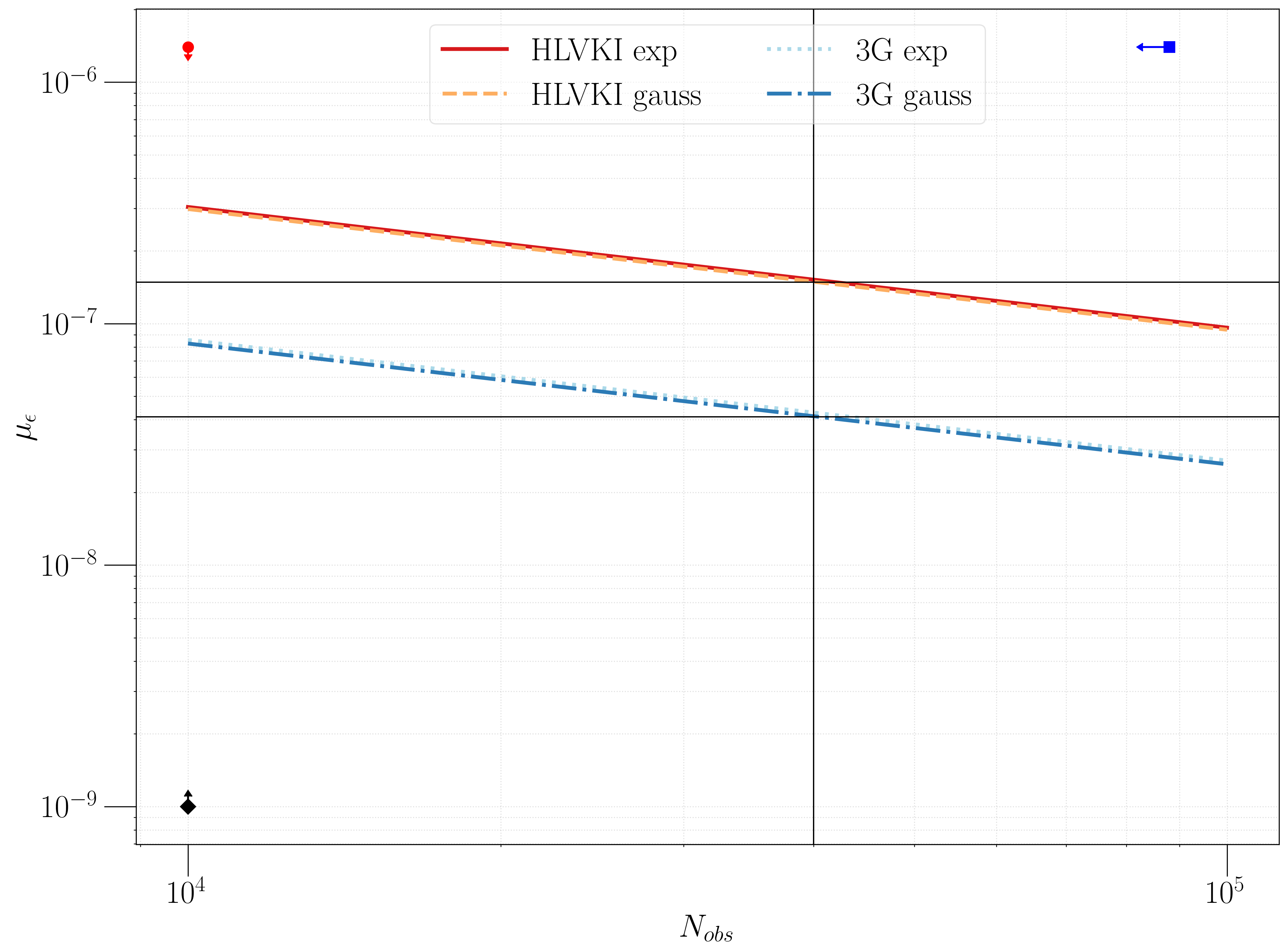}
        \caption{The top panel represents the assumed noise curves for the considered 2G and 3G detectors. In the bottom panel, we show the one-sigma sensitivity of the future detector networks in $N_{\textrm{obs}}-\mu_\epsilon$ plane for the Exponential and Gaussian density profiles for one year of the observational run\footnote{Even though Einstein Telescope (ET) is planned to have three colocated detectors~\cite{ET}, we only considered one detector (ET1) for our test study.}. The horizontal solid lines show the achievable 1-sigma sensitivity for $\mu_\epsilon$ if the in-band pulsars are $N_{\textrm{obs}}=40000$ in total with the 2G and 3G detectors. The arrows with the circle, the square, and the diamond at its end show the $95\%$ confidence upper limits set on $\mu_\epsilon$ and $N_{\textrm{obs}}$ and the predicted lower limit on ellipticity in~\cite{Woan_2018}. }
        \label{fig:future_prediction}
\end{figure} 


\section{Future Sensitivity of the Search}\label{sec:future}
Given the successful completion of the O3 run of Advanced LIGO and Advanced Virgo detectors, the upgrades of second-generation (2G) detectors are planned and aim to achieve the Advanced LIGO Plus (A+) and Advanced Virgo Plus (AdV+) sensitivity during the fifth observing run. Along with the KAGRA detector, situated in Japan (K,~\cite{Kagra1,Kagra2,https://doi.org/10.48550/arxiv.2005.05574}), which has started its operation, they will be joined by the LIGO-India observatory (I,~\cite{ligo_india,Saleem_2021}), which is planned for construction in Hingoli, India with A+ sensitivity. Other than that, the third-generation (3G) observatories such as Cosmic Explorer (CE,~\cite{CE}) and Einstein Telescope (ET,~\cite{ET}) are also envisioned for the future. The design sensitivities of these detectors are shown in top panel of Fig.~\ref{fig:future_prediction} using the publicly available projected noise sensitivity curves~\cite{2G_noise_curves,CE_noise_curves,ET_noise_curves}. 
As the detector network grows and the sensitivities of the detectors improve, it will be interesting to get an idea about the sensitivity of the stochastic targeted search to the parameters of the galactic MSPs population. We measure the sensitivity of the average ellipticity $\mu_\epsilon$
through the expected SNR of the overall amplitude $\langle\rho_{\hat{A}}\rangle$ using Eqs. (\ref{eq:fisher}), (\ref{eq:lkhdDirty}), and (\ref{eq:SNRA}) given as 
\begin{widetext}
\begin{equation}
    \langle\rho_{\hat{A}}\rangle=A \sqrt{\bm{\hat{\mathcal{P}}}^\dagger\mathbf{\Gamma}\bm{\hat{\mathcal{P}}}}=N_{\textrm{obs}}\mu_{\epsilon}^2\left[T_{\mathrm{days}}\sum_{\mathcal{I},t_i,f}\frac{\hat{\mathcal{P}}^*_\alpha\gamma^{\mathcal{I}*}_{\alpha}(t_i;f)\,\gamma^{\mathcal{I}}_{\alpha'}(t_i;f)\hat{\mathcal{P}}_{\alpha'}\,\bar{H}^2_f}{P_1(f)P_2(f)}\right]^{1/2}\,.
\end{equation}
\end{widetext}
Here, we assume that $i$ runs from 0 to the number of segments in one sidereal day, the data is taken for $T_{\mathrm{days}}$ number of sidereal days, and the noise is stationary during the whole observing run. We note that the SNR is proportional to the square root of the number of total segments (or days) and the frequency bins. In the bottom panel of Fig.~\ref{fig:future_prediction}, we show the one-sigma sensitivity (i.e., $\langle\rho_{\hat{A}}\rangle=1$) in $N_{\textrm{obs}}-\mu_\epsilon$ plane for both the exponential and Gaussian density profile considering \textit{one year of observations} with multiple detector network. Here, we have considered a network of five 2G detectors (H, L, V, K, and I) with A+ sensitivity, and for the 3G case, one baseline was formed by assuming one Cosmic Explorer detector in the USA (assuming the location of Hanford detector) and one ET in Europe (assumed the location of Virgo detector).

We note here that, with the 2G detector network with A+ sensitivity, for the optimal number of in-band NSs, $N_{\textrm{obs}}=40000$~\cite{lorimer_2012}, one-sigma sensitivity for $\mu_\epsilon$ is $\sim1.5\times10^{-7}$. Considering the GW detector network with all 2G detectors simultaneously (the HLVKI network) gives only marginal improvement compared to the HL network since the latter favors ORF dominantly. With the 3G detector network, we might achieve $\sim4.1\times10^{-8}$ sensitivity which is close to the minimum limit on the ellipticity~\cite{Woan_2018}. 



\section{Conclusions}\label{sec:conclusions}

We performed a targeted stochastic search for the Galactic millisecond pulsars using the O1, O2, and O3 data from the Advanced LIGO's Hanford $\&$ Livingston and Advanced Virgo detector. In this search, we assumed that the shape of the spectra and spatial distribution of SGWB from the source population is known \textit{a priori} from the theory and the electromagnetic observations. The analysis found that the data is consistent with the noise, favoring the null hypothesis. Hence, we constrained the ensemble properties of the source population, i.e., the in-band number of MSPs, $N_{\textrm{obs}}$ and averaged ellipticity, $\mu_\epsilon$ using the Bayesian formalism. We found that the log-uniform prior for the source properties gives the best $95\%$ confidence upper limits, i.e., $\mu_\epsilon\leq1.4\times10^{-6}$ and $N_{\textrm{obs}}\leq8.8\times10^{4}$. Even though the error bars on our results with the current data are relatively large, we expect them to narrow down with the future network of detectors. We show that with the 3G detectors, we might achieve $\sim4.1\times10^{-8}$ sensitivity which is close to the minimum limit on the ellipticity~\cite{Woan_2018}. 

Meanwhile, many searches have been proposed and performed to set limits on the MSP properties. Recently, matched filtering based targeted search~\cite{https://doi.org/10.48550/arxiv.2111.13106} for the GWs from individual MSP (PSR J0711$-$6830) has set upper limits on the ellipticity $\epsilon\leq 5.3\times10^{-9}$. These searches model the phase evolution of the GW signal given the source parameters, e.g., period, period derivatives, and location in the sky. These searches are more sensitive if the parameter values are known from the electromagnetic observations and the sensitivity degrades considerably for sources with unknown parameters. By performing a hierarchical Bayesian formalism using the GW observation data for known individual pulsars, one can infer the hyperparameters describing the ellipticity distribution (e.g., mean and variance of the ellipticity, if it is Gaussian distributed). In a recent work \citet{PhysRevD.98.063001} use this approach and provides the upper limit for the mean ellipticity using the data from LIGO's sixth science run. The search outlined in our paper complements the matched filtering-based targeted searches. Our method is faster and probes the sources with minimal assumptions for the parameters (i.e., if the only frequency and sky distribution are known). It will be interesting to jointly constrain the ensemble properties using observations from the targeted searches and stochastic searches~\cite{DipongkarNS}. On the other hand, our results are found to be consistent with the upper limits reported in~\cite{PhysRevD.98.063001}. 
Recently,~\citet{DeLillo:2022blw} also inferred the average ellipticity of the Galactic and extragalactic population of the MSPs using the cross-correlation method for SGWBs (as a function of the number of the neutron stars emitting GWs within the frequency band of the search). Given the isotropic background assumptions (this may lead to conservative limits) and the difference in the pulsar population properties, the results are not straightforward to compare with our template-based search. 

It is interesting to note that one can easily extend this work in several directions. One, the actual distribution of the source may differ from the specific spin period, and spatial distribution adapted in our analysis. Hence one can explore the changes in the sensitivity of the search to the variations in source distributions. Second, the uncertainty in the assumed values for the parameter $\langle I^2\rangle$ may play an important role in our analysis. Accounting for this uncertainty can further benefit similar searches in the future. Third, the hyperparameters for the spectral shape, $(\mu,\sigma)$ in Eq.~(\ref{eq:freq_density}) and spatial distribution, $(R_0,z_0/\Sigma_r)$ in Eqs.~(\ref{eq:pdfRadial_exp}) and (\ref{eq:pdfRadial_Gauss}) can be treated as free parameters. Then, one could use the all-sky all-frequency search results~\cite{abbott2021all} to perform a parameter estimation~\cite{O3Iso}. Finally, since the perturbations in the cosmological scale can lead to the anisotropic stochastic background, many models~\cite{anisoAGWB2,anisoAGWB3,anisoAGWB4,anisoAGWB5,anisoAGWB6} can be studied using the formalism discussed in this work.


\begin{acknowledgments}
The authors thank Patrick Meyers for carefully reading the manuscript and providing valuable comments.
This work significantly benefitted from the interactions with the Stochastic Working Group of the LIGO-Virgo-KAGRA Scientific Collaboration. This material is based upon work supported by NSF's LIGO Laboratory, which is a major facility fully funded by the National Science Foundation. The authors are grateful for computational resources provided by the LIGO Laboratory (CIT) supported by National Science Foundation Grants No. PHY-0757058 and No. PHY-0823459, and Inter-University Center for Astronomy and Astrophysics (Sarathi). This research has made use of data or software obtained from the Gravitational Wave Open Science Center (gw-openscience.org), a service of LIGO Laboratory, the LIGO Scientific Collaboration, the Virgo Collaboration, and KAGRA. LIGO Laboratory and Advanced LIGO are funded by the United States National Science Foundation (NSF) as well as the Science and Technology Facilities Council (STFC) of the United Kingdom, the Max-Planck-Society (MPS), and the State of Niedersachsen/Germany for support of the construction of Advanced LIGO and construction and operation of the GEO600 detector. Additional support for Advanced LIGO was provided by the Australian Research Council. Virgo is funded, through the European Gravitational Observatory (EGO), by the French Centre National de Recherche Scientifique (CNRS), the Italian Istituto Nazionale di Fisica Nucleare (INFN) and the Dutch Nikhef, with contributions by institutions from Belgium, Germany, Greece, Hungary, Ireland, Japan, Monaco, Poland, Portugal, Spain. The construction and operation of KAGRA are funded by Ministry of Education, Culture, Sports, Science and Technology (MEXT), and Japan Society for the Promotion of Science (JSPS), National Research Foundation (NRF) and Ministry of Science and ICT (MSIT) in Korea, Academia Sinica (AS) and the Ministry of Science and Technology (MoST) in Taiwan. This article has a LIGO document number LIGO-P2200056.

We used numerous software packages such as {\tt NumPy}~\cite{5725236}, {\tt SciPy}~\cite{Virtanen2020}, {\tt ASTROPY}~\cite{Astropy_2018}, {\tt PyStoch}~\cite{AinPystoch}, {\tt BILBY}~\cite{Ashton_2019,10.1093/mnras/staa2850}, {\tt DYNESTY}~\cite{dynesty}, {\tt PyMultiNest}~\cite{PyMultiNest} and
{\tt MATPLOTLIB}~\cite{Hunter:2007} in this work. We also used the locations information for the detectors (K, I, ET-A) available in the {\tt PyCBC} package~\cite{PhysRevD.90.082004,Usman_2016}.
\end{acknowledgments}

\appendix
\section{Derivation of the PSD for the MSP population}\label{sec:PSD_derivation}
The SGWB can be characterized using a dimensionless energy density parameter which has a unit of Hz$^{-1}$ sr$^{-1}$ defined as~\cite{O3BBR}
 \begin{equation}
     \Omega_\GW{(f,\hatOmega)}=\frac{f}{\rho_c}\frac{\d \rho_\GW}{\d f\d^2\hatOmega}\,,
 \end{equation}
where $\rho_\GW$ is the energy density of the GWs emitted in frequency range $f$ and $f+\d f$ per unit solid angle and $\rho_c=3H_0^2c^2/8\pi G$ is the critical energy density for a flat universe. If, we assume that the SGWB is the resultant of the incoherent sum of GW power from $N(f,\hatOmega)$ number of sources in the frequency range $f$ to $f+\d f$ and lying in solid angle $\hatOmega$ and $\hatOmega+\d^2 \hatOmega$, then 
\begin{equation}\label{eq:OmegaSource}
     \Omega_\GW{(f,\hatOmega)}=\frac{f}{c\rho_c}\sum_{i=1}^{N(f,\hatOmega)}\frac{P^{(i)}}{4\pi r^2_{(i)}}\,\delta(f-f^{(i)})\,\delta^2(\hatOmega-\hatOmega^{(i)})\,.
 \end{equation}
Here $P^{(i)}$ is the GW power radiated from a source at distance $r^{(i)}$ from Earth. Now, the luminosity of the radiated GWs from a deformed axis-symmetric spinning neutron star having moment of inertia $I$ and ellipticity $\epsilon$, emitting nearly monochromatic signal at frequency $f$ is given by~\cite{Maggiore:2007ulw}
 \begin{equation}\label{eq:MSP_power}
     P = \frac{32\,\pi^6\,G}{5c^5}\,\epsilon^2\,I^2\,f^6\,.
 \end{equation}
Considering Eqs.~(\ref{eq:OmegaSource}) and~(\ref{eq:MSP_power}), one can write
\begin{eqnarray}
     \Omega_\GW{(f,\hatOmega)}=\frac{32\,\pi^6\,G\,f}{5c^6\rho_c}\sum_{i=1}^{N(f,\hatOmega)}\frac{\epsilon_{(i)}^2\,I_{(i)}^2\,f_{(i)}^6}{4\pi r^2_{(i)}}\,\nonumber\\
     \times \,\delta(f-f^{(i)})\delta^2(\hatOmega-\hatOmega^{(i)})&&\,,
 \end{eqnarray}
The above expression can be written in terms of population-averaged ($\langle..\rangle_s$) quantities as
\begin{equation}
     \Omega_\GW{(f,\hatOmega)}=\frac{8\,\pi^5\,G}{5c^6\rho_c}N_{\textrm{obs}}\langle \epsilon^2 \rangle_s \,f^7\,p(f)\, \langle I^2 \rangle_{s}\,\langle r^{-2}\rangle_{s}\,p(\hatOmega)\,.
 \end{equation}
Additionally, the dimensionless energy density parameter is related to the source PSD as~\cite{O3BBR}
\begin{equation}
    \Omega_\GW{(f,\hatOmega)}=\frac{2\pi^2}{3H^2_0}f^3 \mathcal{P}(f,\hatOmega)\,.
\end{equation}
Therefore the source PSD can be written as 
\begin{equation}
    \mathcal{P}(f,\hatOmega)=\frac{32\,\pi^4G^2}{5c^8}N_{\textrm{obs}}\langle \epsilon^2 \rangle_s \,f^4\,p(f)\, \langle I^2 \rangle_{s}\,\langle r^{-2}\rangle_{s}\,p(\hatOmega) \,.
\end{equation}

\bibliography{ref.bib}

\end{document}